# Athos - The C# GUI Generator

**Daniela Ilea**
**GM Analytics Software, Timişoara, România**

ABSTRACT. This application comes to help software architects and developers during the long process between user's stories, designing the application's structure and actually coding it.

## 1   General consideration over the application

After spending some time of programming one finds out that the coding is just a small part of your job. Finding what you have to do, the interaction with the client, a good structure of the project provides you about 80% of success in development process. This article is dedicated to the IT specialists willing to become more than just code writers.

The start point of this application came from the misunderstandings between the software architect and the client. We often were in the situation to use UML diagram tools, whose controls look different than Visual Studio's controls. Users looked it, liked it, and after the application were ready: "What? It wasn't supposed to look like that. It looked different, prettier the first time!"

We were in the situation to build a dummy project just to paint the forms and the controls using Visual Studio library, manually add comments to the document... just for a five minutes look from client - a lot of time lost.

In response, we have created an application coming like a bridge between client's needs and developers time, allowing drag drop of the controls (looking pretty much like Visual Studio's controls) in the application, edit properties, comments, export to image and word format. Of course, one might say, there are a lot of UML tools, but the application is doing one more step: also generates the actual C# class, good to be imported in C# application.

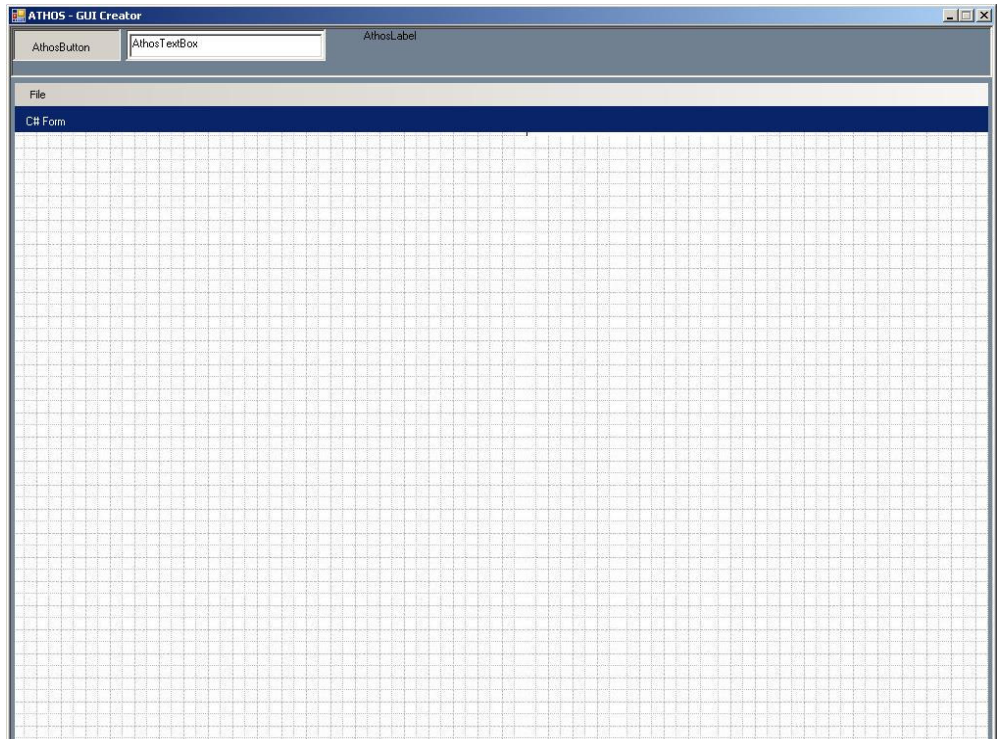

Let us take one step at a time. The user will need some prerequisites before install and use this application:
- administrator rights to save files on disc (or talk to your system administrator to give you rights);
- Microsoft Office 2003 installed;
- .NET framework 2.0 installed (free to download from www.microsoft.com);

The application is using:
- Visual Studio 2005;
- Microsoft Office 2003;
- .NET framework 2.0;

The application is a start point - it uses three controls in this moment: label, text box and button. The flexible architecture allows easily the extension of the application, the possibility to add new controls, new properties for the controls.

All controls are grouped in a separate Class Library project so it can be deployed separately as .dll file or/and attached/used to any other project with different user interface.

The application (a solution like it is called in C#), contains two projects:

- AthosUserControls, containing the actual user controls used in the application, the properties and note dialog
- VisualGUICreator, containing the main form and the interface used to communicate with COM components (Word) and to export into image format. The exporting to Word is an entire article subject, though it will be explained briefly here.

## 2    AthosControl dll

Every control user sees in toolbox is a user control made to look pretty much like C# correspondent control. Every user control has an associated class to communicate with and to get data from. Every user control has a contextual menu attached to get to its properties and to attach a comment to it.

Here we shall illustrate the class diagram (generated using Visual Studio's diagram). You can notice in class diagram the relation between the user control and the class associated to it. Actually, programmatically speaking, the associated class is a property in the user control's class.

Here the application possibilities can be extended. Right now the contextual menu has the options presented as follows.

### 2.1 Properties Dialog

Properties dialog is offered to be extended by Visual Studio, so for the programmers the look will be very familiar. The properties are different from one control to another, although some of them are common: control's name, control's caption, font style (color, size). Here the programmer could add any properties he needs. The property dialog receives this information from the attached class, which is the very same class as the one used to populate the control (Figure 1).

The properties in this class are tagged to appear in Property Box (Figure 2).

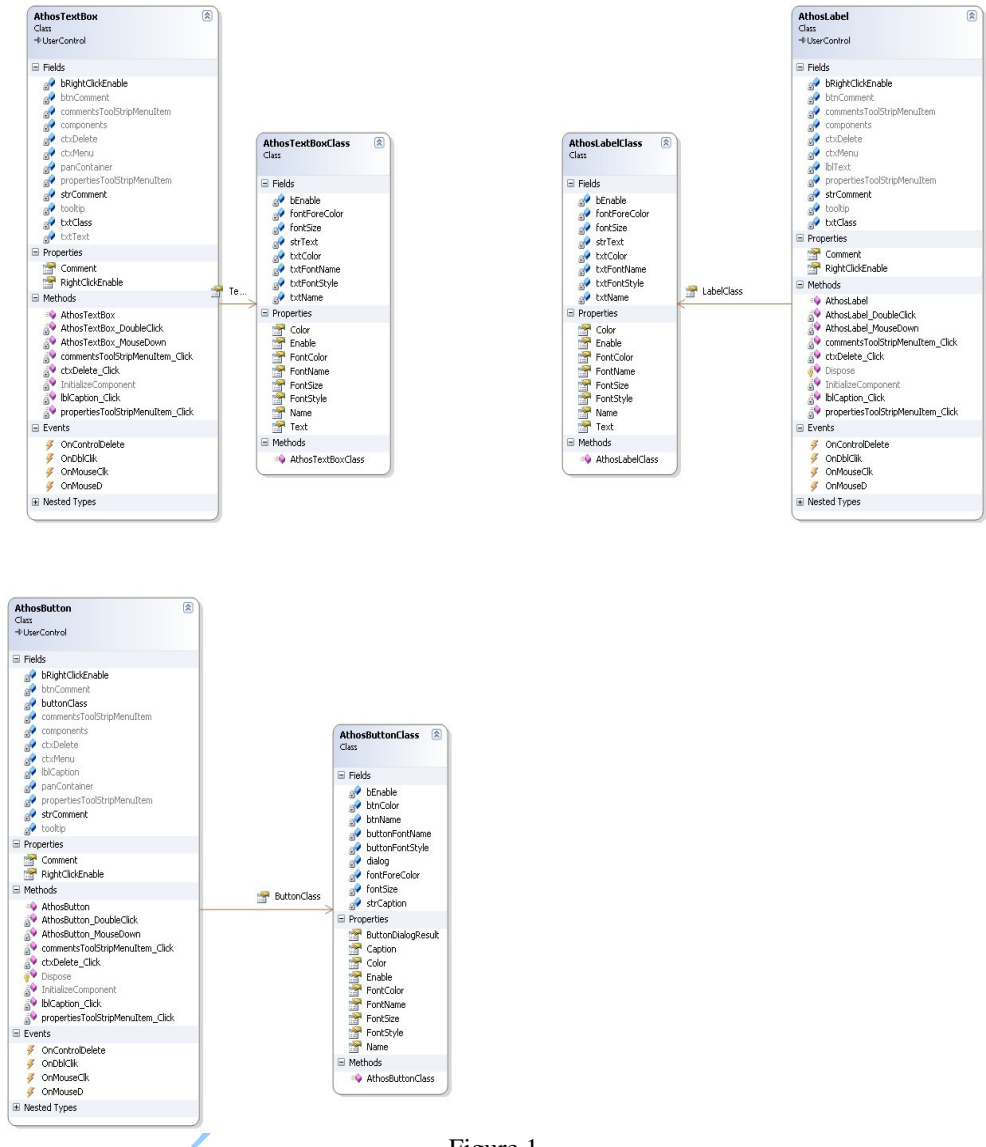

Figure 1

The property box looks like presented in Figure 3.

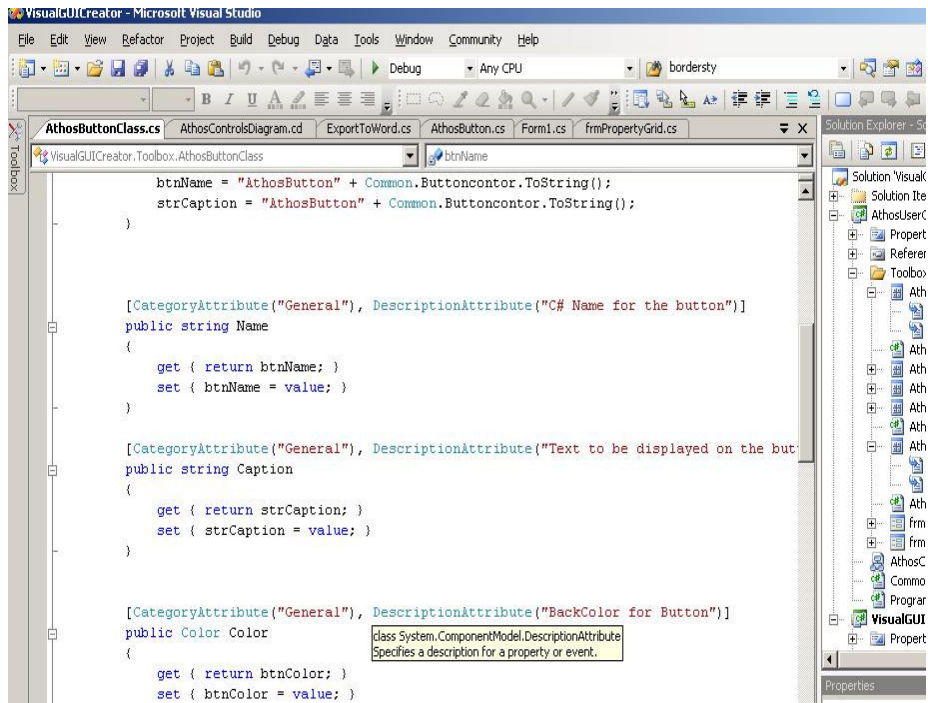
Figure 2

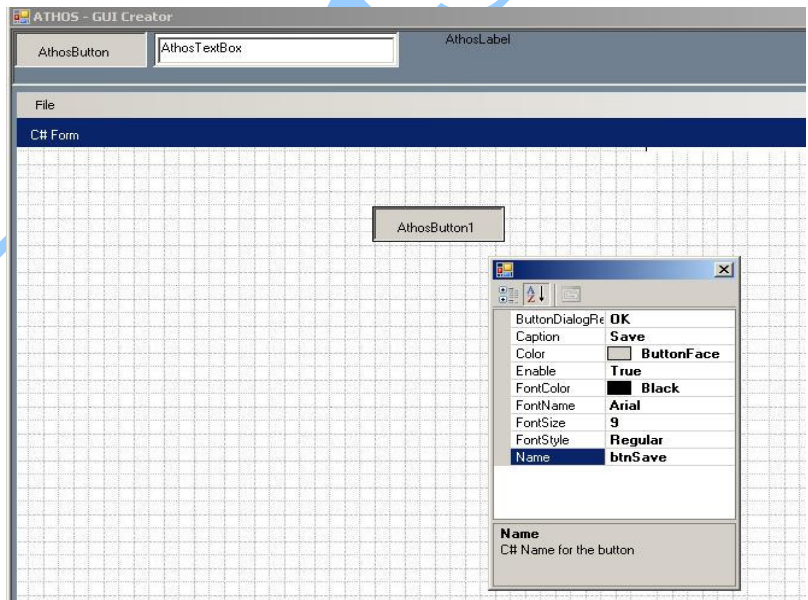
Figure 3

## 2.2 Comment property

Comment property allows the user to make some notes regarding the desired action or content of the control. After a comment is set, the control has a red button telling the user about the comment.

This menu can be extended with other things you might want to attach to a control.

## 3	The main GUI interface

The interface is intended to be as simple as possible. The purpose of this application is obvious to make the programmer's life easier, not to spend much time with it.

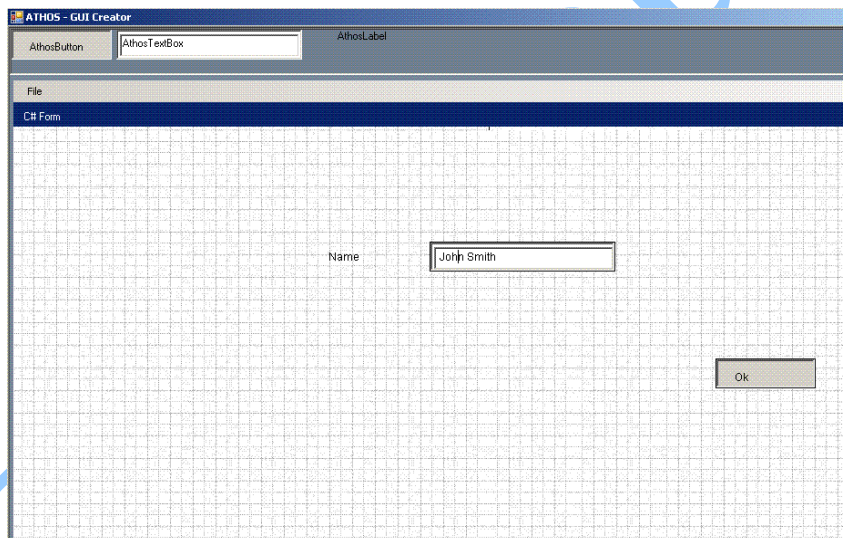

Figure 4

The top part of the application contains the toolbox, the available controls. Below is the work surface. Here the user can add a control and move it in the desired place.

The work surface is actually the future C# form. So, the user has a contextual menu to set the size of the form. After all controls are set, the user can export the exact image of the form in an image file, or, with all details (properties and comments) in a word file. The programmer also can export to C# the form, and a .cs file is generated, ready to be integrated in the project.

In this way the client will be able to see how exactly the application will look like and for the programmer won't be a total lost of time.

## 4   Export to word and image file

To export to Word file we included in solution's references "Microsoft.Interop.Microsoft.Office.Core dll". It exposes the objects needed for export.

```
static Word._Application wordApplication;
static Word._Document wordDoc;
```

A sample of the code has the following format:
```
public  static  void  AddTable(int  rowNumber,  int  colNumber, string
                             strImagePath)
    {
      Word.Range                wordRange                =
wordDoc.Bookmarks.get_Item(ref
                              endOfDoc).Range;
          wordTable = wordDoc.Tables.Add(wordRange,
                      rowNumber,      colNumber,     ref
                      missingValue, ref missingValue);
          wordTable.Range.ParagraphFormat.SpaceAfter = 6;

          wordTable.Rows[1].Range.Font.Bold = 1;
          wordTable.Rows[1].Range.Font.Italic = 1;
    }
```

The word document creates a table and insert in every row and column the characteristics found in the interface. It exports the properties, the comments attached to every control and an image with the document as it looks in the design interface. The design interface has also the possibility to set the dimension of the form. The export to Office subject will be treated in a future work.

## Conclusions

Visual Studio is a very popular and flexible development environment. All programmers should always consider building tools to help their work. Working with Office or image file is very popular in financial world.

Internet provides a lot of open source software, but in the author's opinion, if one can design software, why should use somebody else's? It is more satisfying to develop your own software, adapted to your own needs and extend it if needed and when you need to. To consider developing tools (instead of buying it) is not a waste of time.

C# was the author's choice, and as a five year of .NET development it came natural to develop it. One can use, of course any other programming language.